\documentclass[aps,pre,twocolumn,showpacs]{revtex4-1}  % for review and submission
\usepackage{graphicx}  % needed for figures
\usepackage{dcolumn}   % needed for some tables
\usepackage{bm}        % for math
\usepackage{amssymb,amsmath}   % for math
\usepackage{color}
\renewcommand{\eqref}[1]{(\ref{#1})}

\begin{document}
% the following line is for submission, including submission to the arXiv!!
%\hspace{5.2in} \mbox{Fermilab-Pub-04/xxx-E}

\title{Entropy production in systems with random transition rates}
\author{Daniel M. Busiello, Jorge Hidalgo and Amos Maritan}
\email{maritan@pd.infn.it}
\affiliation{Department of Physics and Astronomy `G. Galilei' and INFN, Universit\'a di Padova, Via Marzolo 8, 35131 Padova, Italy}
\date{\today}

\begin{abstract}
We study the entropy production of a system with a finite number of states connected by random transition rates. The stationary entropy production, driven out of equilibrium both by asymmetric transition rates and by an external probability current, is shown to be composed of two contributions whose exact distributions are calculated in the large system size and close to equilibrium. The first contribution is related to Joule's law for the heat dissipated in a classical electrical circuit whereas the second one has a Gaussian distribution with an extensive average and a finite variance.
\end{abstract}

\maketitle

The second law of thermodynamics states that an isolated system relaxing to equilibrium tends to maximize its entropy.
However, systems out of equilibrium are ubiquitous in nature, being characterized by a continuous exchange of energy or matter with the environment to maintain a non-equilibrium steady state and, thus, producing entropy \cite{jiang-book2004}. For this reason, most of the attempts to develop analogous extremal principles valid for non-equilibrium systems ascribe the entropy production to play the leading role \cite{Gaspar2007}. Some laws have been proposed \cite{jaynes1,jaynes2,martyushev}, stemming from both maximum \cite{dewar} and minimum \cite{prigogine} principles. Nevertheless, the problem remains as one of the greatest challenges of the current statistical physics \cite{goldstein2004,ruelle1997,ruelle2017,reis,gallavotti2004}, partly because of the lack of a universal definition of entropy for systems out of equilibrium \cite{goldstein2004,maes2000,seifert2005}. Indeed, the entropy production has been deeply investigated during the last decade in different contexts, yielding to general fluctuation theorems \cite{bustamante2005,jarzynski2008,seifert2012} and evidencing universal features that are independent on the system details \cite{pigolotti2017}.

We focus on systems amenable to be described by a continuous time Markovian process, whose dynamics follows a Master Equation (ME) \cite{vankampen}. 
In this framework, the coupling with the environment results in a set of stochastic transitions between states, with a net positive entropy production. In particular, for a system of $N$ states with corresponding probabilities $p_i(t)$, $i=1,\dots, N$, at time $t$ and transition rates $w_{i\rightarrow j}$ ($i,j=1,...N$) satisfying the ME
\begin{equation}
\frac{d p_i}{dt} = \sum_{j=1}^N (p_j w_{j\rightarrow i}- p_i w_{i\rightarrow j}),
\end{equation}
the entropy production has been argued by Schnakenberg to be given by \cite{Schnakenberg}:
\begin{equation}
\label{SchnEntProd}
\dot{S} = \sum_{\langle i,j \rangle} (w_{i\rightarrow j} p_i- w_{j\rightarrow i} p_j) \log \frac{w_{i\rightarrow j} p_i}{w_{j\rightarrow i} p_j} 
\end{equation}
where the sum is performed over all pairs of states $i,j$ with non-zero transition rates $w_{i\rightarrow j}$ and $w_{j\rightarrow i}$. 
Note that eq. \eqref{SchnEntProd} is always positive and vanishes for systems where detailed balance holds, $ p_i w_{i\rightarrow j}  = p_jw_{j\rightarrow i}$, and therefore is a good candidate for the entropy production. 
Moreover, eq. \eqref{SchnEntProd} can be derived from the time derivative of a generalized Gibb's entropy with an additional term quantifying the entropy flux exchanged with the environment \cite{Schnakenberg,lebowitz1999,oliveira2012,Amos2015}. Its physical meaning and relation with fluctuation-dissipation theorems have been extensively studied in literature \cite{VandenBroeck1984,maes2003,bustamante2005,oliveira2012,Zia-Schmittmann2007,Gaspar2007,Harris2007,jarzynski2008,seifert2012}.

In his seminal work \cite{Schnakenberg}, Schnakenberg developed a network theory for systems described by a ME, where the states correspond to nodes connected by links representing the transition rates. In this framework, the entropy production can be related to the topological properties of the underlying network (e.g. loops) via closed formulas, where the circulating currents play the most important role \cite{Schnakenberg,Zia-Schmittmann2006, Zia-Schmittmann2007, Gaspar2007, Szabo-Tome-Borsos2010}.

In this Letter we propose a random matrix approach to estimate the entropy production in systems described by a ME. The network does not refer to any particular dynamics, but instead is randomly generated owing to constraints (e.g. system size, connectivity, symmetries in the transition rates) that define an ensemble of systems. Random matrices were first introduced by Wigner in the study of the spectrum of heavy nuclei \cite{wigner1955}, and later have found applications in a wide range of fields \cite{vulpiani2012,allesina2012,sompolinsky1988}. In the same spirit, 
we employ the framework of random matrices to elucidate the interplay between the system dynamics (encoded in the underlying network topology) and the production of entropy.

\begin{figure}[tb]
\centering\includegraphics[width=\columnwidth]{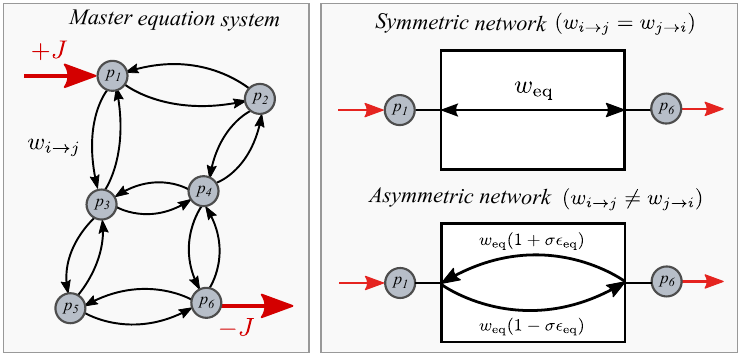}
\caption{(Left) A master equation (ME) system can be represented by a graph in which nodes correspond to states and links to transition rates between them (sketched here as a 6-nodes network). An external probability current $J$ enters into the system by one of the states and exits from another one to ensure stationarity. We analyze the total entropy production when the system is close to equilibrium. (Right) In symmetric ME systems ($w_{i\rightarrow j} = w_{j\rightarrow i}$), the entropy production can be computed from an equivalent network composed exclusively by the external nodes (the ones coupled to the environment via the external current) linked by a symmetric transition rate of strength $w_\mathrm{eq}$. Equivalently, the entropy production of slightly asymmetric networks can be found from an equivalent system composed by the external nodes and two asymmetric links $w_\mathrm{eq}(1\pm\sigma\epsilon_\mathrm{eq})$.
Both $w_\mathrm{eq}$ and $\epsilon_\mathrm{eq}$ encode the topological structure of the underlying network.
}
\label{fig:1}
\end{figure}

We focus on the entropy production of systems at stationarity, $p_i(t)=p_i^*$,  for which eq. \eqref{SchnEntProd} reduces to \cite{oliveira2012}:
\begin{equation}
\label{SchnEntProdSt}
 \dot{S}^* = \sum_{\langle i, j \rangle}(w_{i\rightarrow j} p_i^* - w_{j\rightarrow i} p_j^*) \log \frac{w_{i\rightarrow j}}{w_{j\rightarrow i}}.
\end{equation}
We first consider the simplest case of an open system with symmetric transitions rates, $w_{i\rightarrow j} = w_{j\rightarrow i} = w_{ij}$, which is taken out of equilibrium through the injection of a current of probability $J(t)$ in one of the states. Stationarity is ensured by the ejection of $J(t)$ by another arbitrary state of the system and imposing that $J(t\rightarrow\infty)=J$, i.e. constant at stationarity. 
The system can be mapped into a network as sketched on the left panel of Fig. \ref{fig:1}, 
and the dynamics is described by the ME:
\begin{equation}
 \dot p_i = \sum_{j=1}^N\left(w_{j\rightarrow i} p_j - w_{i\rightarrow j}p_i \right) + J(t) (\delta_{i,1}-\delta_{i,N})
\label{ME}
\end{equation}
for $i=1,...N$, where $w_{i\rightarrow j}=w_{ij}$ in the symmetric case and where, without loss of generality, the current enters in node $1$ and exits through node $N$. The external current $J(t)$ can be mimicked by two \textit{asymmetric} transition rates between nodes $1$ and $N$, $J(t)=p_N(t) \omega_{N\rightarrow 1}(t)-p_1(t) \omega_{1 \rightarrow N}$, where $\omega_{N\rightarrow 1}(t)={(J(t)+\omega_{1\rightarrow N} p_1(t))/}{p_N(t)}$ becomes constant at stationarity.

Writing $\Delta p = p_1 - p_N$, we define the \textit{equivalent transition rate} as $w_\mathrm{eq} = J / \Delta p >0$, which can be understood as a generalized Ohm's law for ME systems. 
Indeed, $w_\mathrm{eq}$ is the strength of a symmetric transition rate in a network composed exclusively by nodes $1$ and $N$ that leads to the same $\Delta p$ for a given current $J$, and depends on the topology of the underlying network (see Fig. \ref{fig:1}). It is also useful to introduce the parameter $\alpha = p_1 + p_N$, so that $p_1 = (\alpha + \Delta p)/2$ and $p_N = (\alpha - \Delta p)/2$. At equilibrium (i.e. when $J=0$), $p_i^*=1/N$, and therefore $\alpha= 2/N$. In terms of the new variables,  
\begin{equation}
 \omega_{N\rightarrow 1} = \omega_{1\rightarrow N} + \frac{2 J \left(w_\mathrm{eq}+ \omega_{1\rightarrow N}\right)}{\alpha w_\mathrm{eq}-J}.
 \label{eq:omegaN1}
\end{equation}

From eq. \eqref{SchnEntProdSt}, it can be seen that internal links do not contribute to the entropy production, and that the only contribution comes from the transition rates between nodes $1$ and $N$:
\begin{equation}
\dot S^* = J \log \left(\frac{\omega_{N\rightarrow 1}}{\omega_{1 \rightarrow N}}\right) \equiv \dot S_\omega. 
\label{eq:Somega}
\end{equation}
We compute the entropy production at stationarity for the system close to equilibrium, i.e. for small values of $J$. Introducing the explicit form of $\omega_{N\rightarrow 1}$ in terms of $J$, eq. \eqref{eq:omegaN1}, and expanding for small values of the current, $J\ll N^{-1}$, we obtain that up to the leading order the entropy production is:
\begin{equation}
\dot S^*  = N \left( \frac{1}{w_\mathrm{eq}} + \frac{1}{\omega_{1\rightarrow N}} \right)  J^2 \equiv  \dot S_\mathcal{J},
\label{eq:S-Joule}
\end{equation}
an expected result since $\dot S^*\geq 0$ and so the linear contribution in $J$ has to be absent.
Eq. (\ref{eq:S-Joule}) consists of two contributions: the first one can be viewed as the entropy production of the system itself, while the second term stems from dissipations in the environment \cite{Schnakenberg}. The form of $\dot S_\mathcal{J}$ (eq. \eqref{eq:S-Joule}) constitutes a fluctuation-dissipation type relation \cite{kubo1966,baiesi} and it resembles Joule's law for the heat dissipated in a classical electrical circuit with equivalent conductance $w_\mathrm{eq}/N$, taking also into account the heat dissipated by the battery. The ideal battery corresponds to the limit case $\omega_{1\rightarrow N}= \infty$.

Using the parallelism with the electrical circuit, the stationary state $p_i^*$ can be retrieved from a variational principle as the one that minimizes the entropy production, with the constraint imposed by the external flux. Such a parallelism also works in the opposite way, as any electrical circuit composed exclusively by multiple resistors can be mapped into a ME system, and its corresponding entropy production turns out to be proportional to the heat dissipation rate (see SI).

\begin{figure}[tb]
\centering
\includegraphics[width=\columnwidth]{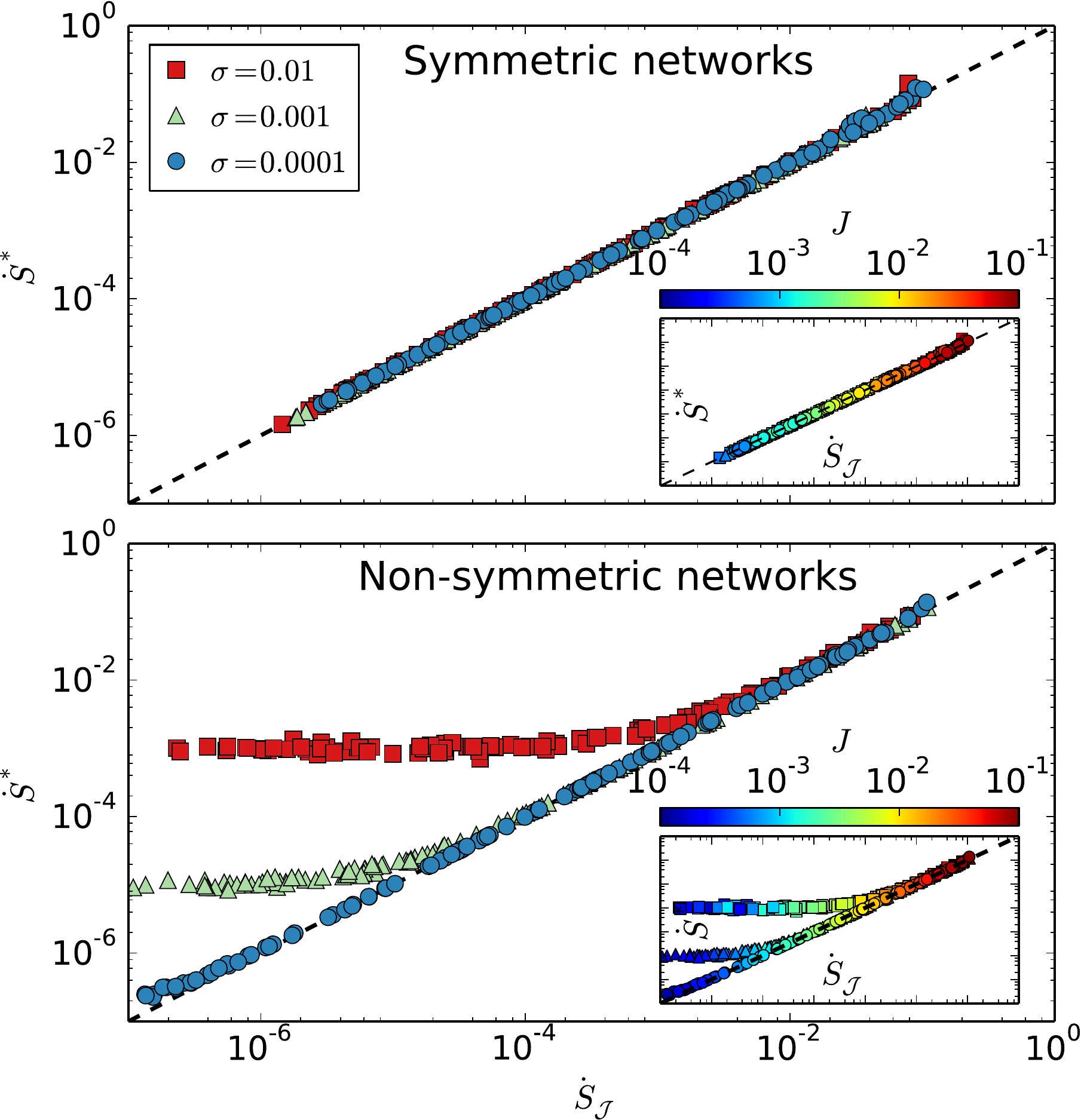}
\caption{\textbf{Entropy production of ME systems represented by random (Erdos-Renyi) networks, compared to the theoretical prediction given by Eq. \eqref{eq:S-Joule}}. Transition rates $w_{i\rightarrow j}$ are randomly generated from a Gaussian distribution with mean $w=1$ and standard deviation $w\sigma$. Each dataset is composed by $100$ realizations of networks of size $N=25$, fixing the parameter $\omega_{1\rightarrow N}=10$. The external flux $J$ is a random variable, being $\log_{10} J$ uniformly distributed in $[-4, -1]$ (so that $J\in[10^{-4},10^{-1}]$). All plots are in log-log scale. \textbf{(Top panel)} Symmetric networks, for which $w_{i\rightarrow j} = w_{j\rightarrow i}$ for each pair of links. The connectivity $K$ is a uniform random variable in $K\in[0.25, 1]$. \textbf{(Bottom Panel)} - Non-symmetric networks: we remove the symmetric constraint, and therefore $w_{i\rightarrow j}\neq w_{j\rightarrow i}$ in general. In this case, we have fixed the connectivity $K=0.5$. In both panels, insets represent the same corresponding points but using a color scale for the external probability flux $J$. Similar results can be obtained for other network topologies different from the Erdos-Renyi (Eq. \eqref{eq:S-Joule} is valid in general).}
\label{fig:2} 
\end{figure}

Before moving to the more general case of  non-symmetric transition rates and also for later comparison reasons we have checked numerically the validity of eq. \eqref{eq:S-Joule} on randomly generated ME systems of size $N$ and connectivity $K$ (defined as the fraction of non-directed connections respect to the $N(N-1)/2$ links in the fully-connected case).
In the symmetric case, each non-null entry of $w_{i\rightarrow j}$ is taken from a Gaussian distribution with average $w$ and standard deviation $w\sigma$, with the constraint that $w_{i\rightarrow j} = w_{j\rightarrow i}$ for each pair of links. 
We numerically integrate the corresponding ME and compute the entropy production at stationarity. The result is compared with eq. \eqref{eq:S-Joule}, where all the dependency on the network topology in eq. \eqref{eq:S-Joule} is encapsulated in the parameter $w_\mathrm{eq}$. Results are shown in Top panel of Fig. \ref{fig:2} (in log-log scale), evidencing that eq. \eqref{eq:S-Joule} perfectly works for a wide range of values of $J$.

Now we want to generalize the simple result given by eq. \eqref{eq:S-Joule} to  networks generated as in the symmetric case but taking $w_{i\rightarrow j}$ and $w_{j\rightarrow i}$ as independent random variables. For each topology, $w_\mathrm{eq}$ is simply estimated taking $w_{i\rightarrow j}=w$. Then, we compute the entropy production of each network in the presence of a probability current $J$, and compare its value with the one given by $\dot{S}_{\mathcal{J}}$. Results are presented in the Bottom panel of Fig. \ref{fig:2} for different values of the heterogeneity $\sigma$; the prediction given by eq. \eqref{eq:S-Joule} fails in the limit of small flux, as detailed balance does not holds for a general asymmetric network, and therefore $\dot S \neq 0$ when $J\rightarrow 0$. 

Deviations from eq. \eqref{eq:S-Joule} due to asymmetries in the network can be described within a perturbative framework when the system is close to equilibrium. 
We focus on ensembles of random Erdos-Renyi topologies of connectivity $K$ for which the non-null elements of $w_{i\rightarrow j}$ are independent Gaussian random variables of mean $w$ and standard deviation $w\sigma$. 
Erdos-Renyi topologies are good proxies for networks with limited connectivity for which the property of small world holds \cite{newman2003}. Notice that, in the (rather general) case of a system composed by $n$ elementary units, each of them taking $m$ possible values, the total number of states is $N=n^m$. If the dynamics allows that each unit can change its value once at a time, the system can pass from one state to another through, roughly, $n$ individual steps. This leads to a typical path distance of the order of $n=\log_m(N)$, which is actually the small world condition \cite{newman2003}.
Introducing the adjacency matrix $A_{ij}=\Theta[w_{i\rightarrow j}]$ ($\Theta$ is the Heaviside step function with $\Theta(0)=0$), we can write $w_{i\rightarrow j}=w A_{ij} (1+\sigma \epsilon_{ij})$ in terms of independent Gaussian variables $\epsilon_{ij}$ of zero mean and unit variance. 

The entropy production given by eq. \eqref{SchnEntProdSt} can be split into two contributions, 
\begin{equation}
\dot S^* = \dot S_\mathrm{int} + \dot S_\omega, 
\end{equation}
where $\dot S_\mathrm{int}$ is the one given by internal links in the network and $\dot S_\omega$ the one given by the external links connecting nodes $1$ and $N$, eq. \eqref{eq:Somega}. Formally, such a separation can be done if we do not allow the presence of internal links $w_{1\rightarrow N}$ and $w_{N\rightarrow 1}$, although whether considering them or not does not significantly change the value of the entropy production in large networks.

We want to show that close to equilibrium, $J,\sigma \ll N^{-1}$, the entropy production, $\dot S^*$, differs from the corresponding one in absence of asymmetry, $\dot S_\mathcal{J}$, as given by eq.(\ref{eq:S-Joule}), by a gaussian random variable of mean $(KN-2-K + \mathcal{O}(1/N))w\sigma^2$ and variance $4Kw^2\sigma^4 + \mathcal{O}(1/N)$.
To show that we write the stationary state as $p_i^* \simeq \left(1 + \sigma q_i + J r_i \right)/N$; the normalization constraint implies that $\sum_i q_i = \sum_{i} r_i = 0$. Then, we expand the entropy production in terms of $\sigma$ and $J$, neglecting contributions of order higher than $\sigma J N^2$, $\sigma^2 N^2$ and $J^2 N^2$.  Defining the antisymmetric matrix $D_{ij}=\epsilon_{ij} - \epsilon_{ji}$ (whose entries are Gaussian variables with zero mean and standard deviation $\sqrt{2}$), the first order contributions to $\dot S_\mathrm{int}$ are:
\begin{equation}
\frac{\dot S_\mathrm{int}}{w\sigma^2} = \frac{1}{N} \sum_{\langle i,j \rangle}A_{ij} \left(D_{ij}^2 + D_{ij}\left(q_i + \frac{J}{\sigma} r_i \right)\right).
\label{eq:Ssigma1}
\end{equation}
For large $N$, the first term of the r.h.s. of eq. \eqref{eq:Ssigma1}, becomes a Gaussian random variable of mean $K(N-1)$ and standard deviation $2 \sqrt{K}$ by means of the central limit theorem (see SI). Remarkably, this term is positive and scales linearly with the system size $N$. 
The second and third terms in eq. \eqref{eq:Ssigma1} require to solve the ME close to equilibrium to obtain the values of $q_i$ and $r_i$. Up to first order in $N$, we find that $q_i \simeq - \frac{1}{K N} \sum_{j} A_{ij} D_{ij}$ (see SI). Such a dependency is meaningful: if in a node $i$, most of $D_{ij}=(w_{i\rightarrow j}-w_{j\rightarrow i})/w>0$ (resp. $<0$), an outgoing (incoming) flux is expected between node $i$ and its neighbors, and therefore $p_i^* < 1/N$ ($>1/N$). 
Introduced in eq. \eqref{eq:Ssigma1}, the second term gives a Gaussian random contribution to $\dot S_\mathrm{int}$ with mean $-2$ and standard deviation $\sqrt{8/N}$ (see SI).
Equivalently, solving the ME we find that, up to first order in $N$, $r_i \simeq (\delta_{i1}-\delta_{iN})/(wK)$ (see SI), evidencing the major role of the external nodes. This term gives another Gaussian contribution to $\dot S_\mathrm{int}$ which has zero mean and standard deviation $(2/\sqrt{KN})J/\sigma$ (see SI). Assembling all the terms together, the leading contributions give that $\dot{S}_\mathrm{int}$ is normally distributed with mean  and  variance given by:
\begin{eqnarray}
\label{eq:gaussian1}
\langle \dot{S}_\mathrm{int} \rangle = \left( K N - (2+K) \right) w \sigma^2 +\mathcal{O}(1/N)\\
\label{eq:gaussian2}
\langle (\dot{S}_\mathrm{int})^2 \rangle -  \langle \dot{S}_\mathrm{int} \rangle^2= 4Kw^2\sigma^4 + \mathcal{O}(1/N)
\end{eqnarray}

\begin{figure}[t]
\centering
\includegraphics[width=\columnwidth]{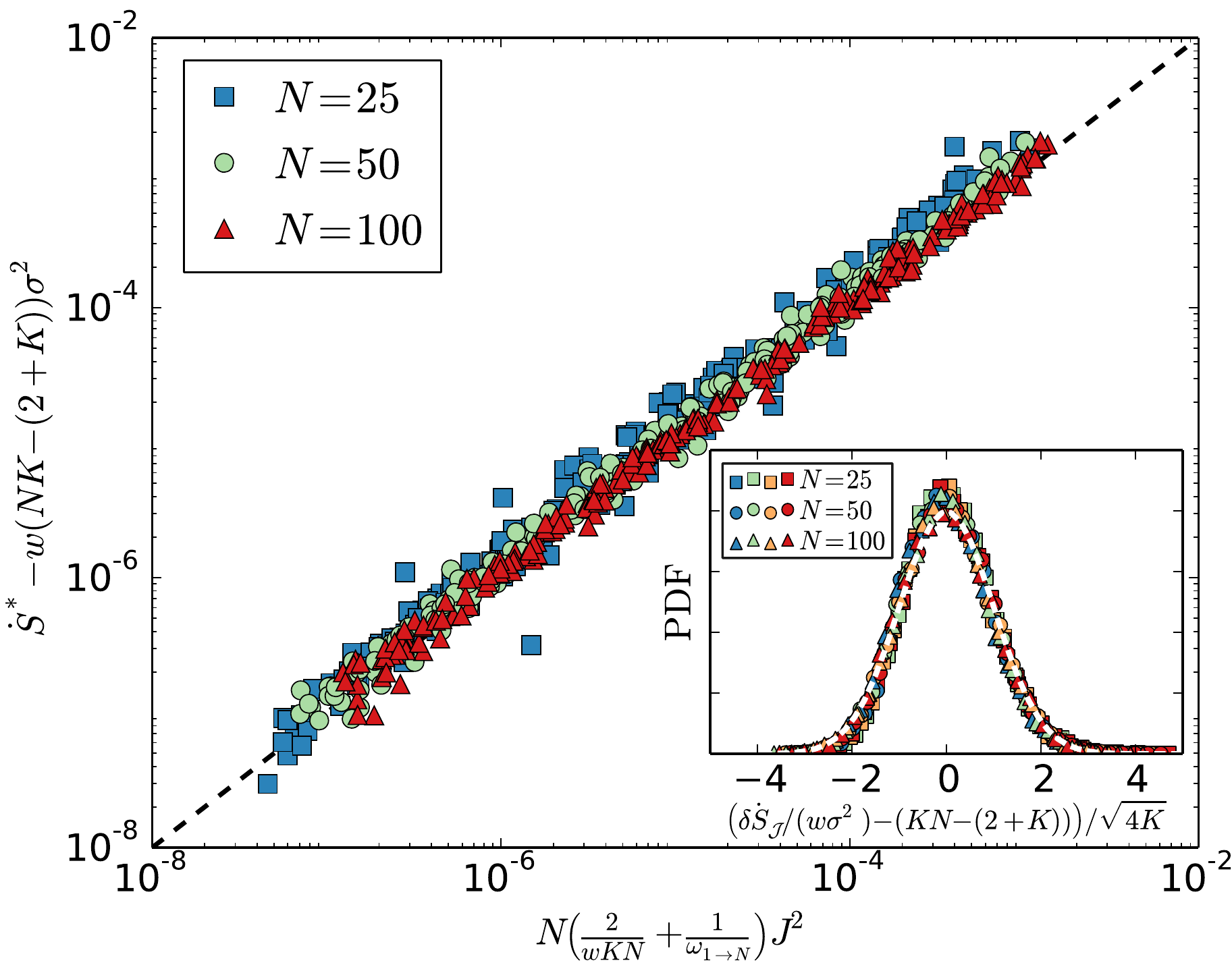}
\caption{\textbf{Deviations from eq. \eqref{eq:S-Joule} follow generic scaling relations}. The relation between $\dot{S}^{*}$ and $J$ for networks generated with different $N$, $K$, $J$ and $\sigma$ can be enlightened plotting $\dot{S}^{*}-\langle\delta\dot{S}_\mathcal{J}\rangle$ against $\langle\dot{S}_\mathcal{J}\rangle$. All datasets collapse into the straight line $y=x$ (with deviations around). We have generated 250 networks for each system size with a random connectivity uniformly distributed in $K\in[0.25,1]$ and a external flux $J$ such that $\log_{10}J$ is a uniform random variable in $[-4,-1]$ ($J\in[10^{-4},10^{-1}]$); the heterogeneity has been taken $\sigma=J$ for each realization (this ensures that the current and heterogeneity contributions to the entropy production are of the same order and therefore both have to be considered). (Inset) We collapse the PDFs of $\delta\dot{S}_\mathcal{J}$ by subtracting the contribution given by Joule's law, eq. \eqref{eq:S-Joule}, and the mean $w(KN-(2+K))\sigma^2$, and dividing by the standard deviation $w\sqrt{4K}\sigma^2$ (blue, green, yellow and red points correspond to $K=0.25,0.5,0.75$ and $1$, respectively). The corresponding $w_\mathrm{eq}$ is computed setting $w_{i\rightarrow j}=w$ for each network topology. All histograms collapse into a Gaussian distribution of zero mean and unit variance (dashed line). We have generated $10^4$ independent realizations for each set of parameters, setting $J=\sigma=10^{-3}$.
We have set $\omega_{1\rightarrow N}=10$ in both panels.
}
\label{fig:3} 
\end{figure}

On the other hand, the term $\dot S_\omega$ can be computed in terms of a generalized Ohm's law for non-symmetric networks close to equilibrium, $\Delta p = J/w_\mathrm{eq} + \sigma \epsilon_\mathrm{eq}$, where  $1/w_\mathrm{eq} = (r_1-r_N)/N$ and  $\epsilon_\mathrm{eq}=(q_1-q_N)/N$ is a network parameter accounting for the asymmetry-induced unbalance between nodes $1$ and $N$. More precisely, $\epsilon_\mathrm{eq}$ represents the asymmetry of an equivalent network composed exclusively by nodes $1$ and $N$  with the same $\Delta p$ and  $(w_{N\rightarrow 1}-w_{1\rightarrow N})/(w_{N\rightarrow 1}+w_{1\rightarrow N})=\sigma\epsilon_\mathrm{eq}$ (see Fig. \ref{fig:1}). Notice that $w_\mathrm{eq}$ and $\epsilon_\mathrm{eq}$ exclusively depends on the network structure. In terms of these parameters, the leading contributions to $\dot S_\omega$ are:
\begin{equation}
 \dot S_\omega(\sigma) = \dot S_\mathcal{J} + N \epsilon_\mathrm{eq} \sigma J,
 \label{eq:Somega-sigma}
\end{equation}
where $\dot S_\mathcal{J}$ is defined in eq. \eqref{eq:S-Joule}. Deviations from eq. \eqref{eq:S-Joule}, $\delta \dot S_\mathcal{J}=\dot S^* - \dot S_\mathcal{J}$ are given by $\delta \dot S_\mathcal{J}=\dot{S}_\mathrm{int} + N \epsilon_\mathrm{eq} \sigma J$. For large networks we found that $\epsilon_\mathrm{eq}$ is normally distributed, 
$\epsilon_\mathrm{eq} \sim \mathcal{N}\left(0, 2/\sqrt{KN^3}  \right)$ (see SI). Consequently, deviations from eq. \eqref{eq:S-Joule} are essentially given by 
$\delta \dot S_\mathcal{J}\approx\dot{S}_\mathrm{int}$, which is our main result.

The relation between $S^*$ and $J$ can be enlightened representing $\dot{S}^*-\langle \delta \dot S_\mathcal{J} \rangle = \dot{S}^*-(N K -(2+K)) w \sigma^2$ against $\langle \dot S_\mathcal{J} \rangle = N\left(\langle w_\mathrm{eq}^{-1} \rangle + \omega_{1\rightarrow N}^{-1}\right)J^2$. We can compute the distribution of $w_\mathrm{eq}$ for an ensemble of random networks of size $N$ and connectivity $K$; interestingly, this problem is equivalent to computing the equivalent conductance of an electrical circuit in which resistors are randomly connected. $P(w_\mathrm{eq})$ becomes a Gaussian distribution with mean $NKw/2$ and standard deviation $w\sqrt{NK(1-K)/8}$ for large networks (see SI), becoming narrower when $N$ increases (relatively to its mean). Therefore, we can use the approximation $\langle w^{-1}_\mathrm{eq} \rangle \sim eq 2/(NKw)$ (we refer to the SI for a more formal derivation). Pairs of $(S^*,J)$ are represented with this procedure  in the main panel of Fig. \ref{fig:3}, evidencing that the entropy production of non-symmetric ME systems of different sizes $N$, connectivity $K$, heterogeneity $\sigma$ and external current $J$, all collapse around the straight line $y=x$, with deviations around decreasing for larger systems.

Finally, we check numerically that deviations from eq. \eqref{eq:S-Joule} follow the scaling relations given by equations \eqref{eq:gaussian1} and \eqref{eq:gaussian2}. 
Subtracting the mean and dividing by the standard deviation, all the histograms collapse into the Gaussian distribution with zero mean and unit variance, as depicted in the inset of Fig. \ref{fig:3}.

Summarizing, in this Letter we have employed the network theory proposed by Schnakenberg to analyse the entropy production of systems out of equilibrium from a random matrix perspective. We have obtained general scaling relations for the entropy production of network ensembles representing a set of systems and dynamics, and we have identified the topological parameters ($w_\mathrm{eq}$ and $\epsilon_\mathrm{eq}$) that play the most important role.

Our framework provides a null-model on which compare the entropy production of specific systems and dynamics. In particular, some biological systems seem to have evolved in order to dissipate energy and produce entropy at the minimum or maximum possible rate \cite{martyushev,kleidon2010}. It could be interesting to analyse the topological features of real dynamics (in particular the values of $w_\mathrm{eq}$ and $\epsilon_\mathrm{eq}$) with what we have obtained for random networks, which have not been obtained through optimization/evolution processes.

Finally, our results has been derived for small world-like networks with an homogeneous degree distribution of finite variance. Interesting generalizations include cases where the degree distribution has not a finite variance, e.g. scale free-networks \cite{barabasi1999}, strong asymmetric transition rates and transition rates between different pairs of states with non-trivial correlations.

We thank S. Suweis and S. Azaele for useful discussions and suggestions.

%\bibliographystyle{plainnat}
%\bibliography{Draft_v12_PRL_biblio.bib}


\newpage
\widetext
\newpage
\appendix
\renewcommand{\thepage}{S\arabic{page}}
\setcounter{page}{1}

\begin{center}
{\large\bf Supplementary Information for}\\
{\Large\bf Entropy production in systems with random transition rates}\vspace*{0.5cm}\\
D. M. Busiello, J. Hidalgo and A. Maritan
\end{center}

\section{Mapping: from an electrical circuit into a ME system}
\label{sec:circuit}
Let us consider a circuit $\mathcal{C}$ composed of $N$ nodes connected by multiple resistors, with conductances equal to $c_{ij}$. When a difference of potential $\Delta V = V_1-V_N$ is applied between nodes $1$ and $N$, the potential at each node $i$, $V_i$, is given by Kirchoff's law, which ensures the minimization of the heat dissipation:
\begin{equation}
\dot Q = \sum_{\langle i,j \rangle} c_{ij} (V_i-V_j)^2
\label{heat}
\end{equation}
where the sum is performed over all pairs of nodes $i$, $j$. 
One can then build a fictitious dynamics for the potentials, whose stationary state corresponds to the physical state of $\mathcal{C}$. To proceed, we fix $V_1$ and $V_N$ (with $V_1-V_N=\Delta V$), and consider the following dynamics for nodes $i=2,...,N-1$:
\begin{equation}
\tau \dot V_i(t) = -\frac{\partial \dot Q}{\partial V_i}= \sum_{j=1}^N c_{ij} \left( V_j - V_i \right).
\label{MEcirc}
\end{equation}
An arbitrary time scale $\tau$ can be introduced without changing the stationary state. Notice that eq. \eqref{MEcirc} leads to the state which minimizes $\dot Q$, as $\ddot Q(t) = \sum_{i} \frac{\partial \dot Q}{\partial V_i} \dot V_i = - 2\tau\sum_i \dot V_i^2 < 0$.
Let us notice that, despite its analogy, eq. \eqref{MEcirc} is not a master equation, since $\sum_{i=1}^N \dot V_i(t)\neq0$ for arbitrary $t$, and therefore normalization is not conserved.  However, one can relax the constraint on nodes $1$ and $N$ to be obeyed only at stationarity. Defining the normalized potentials, $v_i=V_i/\sum_{i}V_i=V_i/V_T$ and introducing the new parameters $c_{1\rightarrow N}$ and $c_{N\rightarrow 1}$ ($c_{1\rightarrow N}\neq c_{N\rightarrow 1}$ in general), we study the following ME dynamics:
\begin{equation}
\tau \dot v_i(t) = \sum_{j}c_{ij}(v_j-v_i) + \left(c_{N\rightarrow 1}v_N - c_{1\rightarrow N} v_1\right) (\delta_{i,1} - \delta_{i,N})
\label{MEcirc2}
\end{equation}
for $i=1,...,N$, where $c_{N\rightarrow 1}$ and $c_{1\rightarrow N}$ have to be chosen to ensure $v_1-v_N=\Delta V/V_T$ at stationarity. 
Notice that $\sum_{i=1}^N \dot v_i = 0$ for arbitrary $t$ in eq. \eqref{MEcirc2}, which has an equivalent form to eq. \eqref{ME}. In addition, it leads to the state minimizing the heat dissipated by the circuit.
Without loss of generality we can fix $\tau=1$.

The normalized current, $J = c_{N\rightarrow 1}v_N - c_{1\rightarrow N} v_1 = c_\mathrm{eq} \Delta V/V_T = I/V_T$, where $c_\mathrm{eq}$ is the conductance of the equivalent circuit formed by one single resistor and $I$ is the supplied electrical current. For small values of the external current, Schnakenberg's entropy production at stationarity is given by eq. \eqref{eq:S-Joule}:
\begin{equation}
\label{SchnLimit0}
\dot{S}^*= N\left(\frac{1}{c_\mathrm{eq}}+\frac{1}{c_{1\rightarrow N}}\right) J^2,
\end{equation}
which corresponds to Eq. \eqref{eq:S-Joule} of the main text with $c_\mathrm{eq}= w_\mathrm{eq}$ and $c_{1\rightarrow N}=\omega_{1\rightarrow N}$. Written in terms of the heat dissipated by the circuit, $\dot Q = I^2/c_\mathrm{eq}$, the entropy production of the associated ME dynamics is:
\begin{equation}
\label{SchnLimit}
\dot{S}^* = \frac{N}{V_T^2} \left(1 + \frac{c_\mathrm{eq}}{c_{1\rightarrow N}} \right) \dot Q.
\end{equation}
Its important to note that the relation given by eq. \eqref{SchnLimit} is determined up to the specific choice of parameters $V_T$ and $c_{1\rightarrow N}$, as this is an intrinsic ambiguity of the mapping. Furthermore, there is a dissipation contribution coming from the battery, which vanishes in the the limit of a ``perfect'' battery, $c_{1\rightarrow N}\rightarrow \infty$.

\section{Calculation of $\dot S_\mathrm{int}$}
Writing the transition rates $w_{i\rightarrow j}$ in terms of independent Gaussian variables $\epsilon_{ij}$ of zero mean and unit variance, $w_{i\rightarrow j}=A_{ij} w (1+\sigma \epsilon_{ij})$ (with $A_{ij}=\Theta[w_{i\rightarrow j}]$) and introducing $p_i^*=(1+\sigma q_i + J r_i)/N$ in eq. \eqref{SchnEntProdSt}, we obtain:
\begin{equation}
 \dot S_\mathrm{int} = \sum_{\langle i,j \rangle} A_{ij} w \left( 
 (1+\sigma \epsilon_{ij}) \left(\frac{1}{N}\left(1 + \sigma q_i + J r_i \right)\right) - 
 (1+\sigma \epsilon_{ji}) \left(\frac{1}{N}\left(1 + \sigma q_j + J r_j \right)\right)
 \right) \log \left(\frac{1+\sigma \epsilon_{ij}}{1+\sigma \epsilon_{ji}}\right).
\end{equation}
Close to equilibrium ($\sigma\ll N^{-1}, J\ll N^{-1}$), the leading contributions are:
\begin{eqnarray}
 \label{eq:Sint1}
 \dot S_\mathrm{int}^1 & = & \frac{w \sigma^2}{N} \sum_{\langle i,j \rangle} A_{ij} D_{ij}^2\\
 \label{eq:Sint2}
 \dot S_\mathrm{int}^2 & = & \frac{w \sigma^2}{N} \sum_{\langle i,j \rangle} A_{ij} D_{ij} (q_i-q_j) = \frac{w \sigma^2}{N} \sum_{i,j} A_{ij} D_{ij} q_i \\
 \label{eq:Sint3}
 \dot S_\mathrm{int}^3 & = & \frac{w \sigma J}{N} \sum_{\langle i,j \rangle} A_{ij} D_{ij} (r_i-r_j) = \frac{w \sigma J}{N} \sum_{i,j} A_{ij} D_{ij} r_i,
\end{eqnarray}
where $D_{ij}=\epsilon_{ij}-\epsilon_{ji}$. In the next section we compute each term separately, and we show that the major contribution is given by $\dot S_\mathrm{int}^1$, whereas the second and the third term can be neglected for large network sizes. For convenience, we will call $\mathcal{N}(a,b)$ the Gaussian distribution with mean $a$ and standard deviation $b$.

It is worth mentioning that, in our calculations, we do not distinguish the cases in which the external nodes are or not connected also by internal nodes ($A_{1N}=0$ and $1$ respectively), as this constitutes a minor contribution for the total entropy production.

\subsection{Calculation of $\dot S_\mathrm{int}^1$}
For each pair of links $i$ and $j$, $D_{ij}$ is an independent random variable of zero mean and standard deviation $\sqrt{2}$. Therefore, $D_{ij}^2$ is distributed as a chi-square distribution whose mean and variance can be easily calculated:
\begin{eqnarray}
 \langle D_{ij}^2 \rangle &=& 2 \\
 \langle (D_{ij}^2-\langle D_{ij}^2\rangle)^2\rangle &=& \langle D_{ij}^4 \rangle - \langle D_{ij}^2 \rangle^2 = 2 \langle D_{ij}^2 \rangle ^2 = 8.
\end{eqnarray}
Eq. \eqref{eq:Sint1} involves the sum of $K N (N-1)/2$ independent variaeq.bles of this type. When $N$ is large, by means of the Central Limit Theorem, we obtain that $\dot S_\mathrm{int}^1$ follows a normal distribution:
\begin{equation}
 P\left(\frac{\dot S_\mathrm{int}^1}{w \sigma^2}\right)  = \mathcal{N}\left(K(N-1), 2\sqrt{K\left(1-\frac{1}{N}\right)}\right) \simeq \mathcal{N}(K (N-1), 2\sqrt{K}).
 \label{eq:P-Sint1}
\end{equation}

\subsection{Calculation of $\dot S_\mathrm{int}^2$}
In the absence of external flux ($J=0$) and for small asymmetry $\sigma$, the stationary state can be written as $p_i^* = \frac{1}{N}(1+\sigma q_i)$. The values of $q_i$ can be retrieved solving the master equation at stationarity:
\begin{equation}
0 = \sum_{j} A_{ij}w\left((1+\sigma\epsilon_{ji})\left(\frac{1}{N}\left(1+\sigma q_j\right)\right) - (1+\sigma\epsilon_{ij})\left(\frac{1}{N}\left(1+\sigma q_i\right)\right)\right),
\end{equation}
that, up to first order in $\sigma$, leads to the implicit solution:
\begin{equation}
q_i = \frac{1}{k_i}\left( -\sum_j A_{ij} D_{ij} + \sum_j A_{ij} q_j  \right),
 \label{eq:qi}
\end{equation}
where $k_i=\sum_j A_{ij}$ is the number of nodes connected to node $i$. For convenience, we introduce self-interactions ($A_{ii}=1$), as this choice does not change the dynamics in the master equation and leads to simpler expressions in our notation. In the fully connected network ($A_{ij}=1$), the second term on the r.h.s. of eq. \eqref{eq:qi} vanishes (as $\sum_j q_j = 0$), leading to the closed form $q_i = -\frac{1}{N} \sum_j D_{ij}$. This result is rather intuitive: if, for instance, the average rate of the outgoing links in a node $i$ is greater than the ingoing one (the same argument applies in the opposite case) $\sum_j\epsilon_{ij}>\sum_j \epsilon_{ji} \Longrightarrow \sum_{ij}D_{ij}>0$, there is a net outgoing flux to its neighbors and therefore $q_i<0$, or equivalently, $p_i^* < 1/N$.

\begin{figure}[h]
 \centering
 \includegraphics[width=0.5\textwidth]{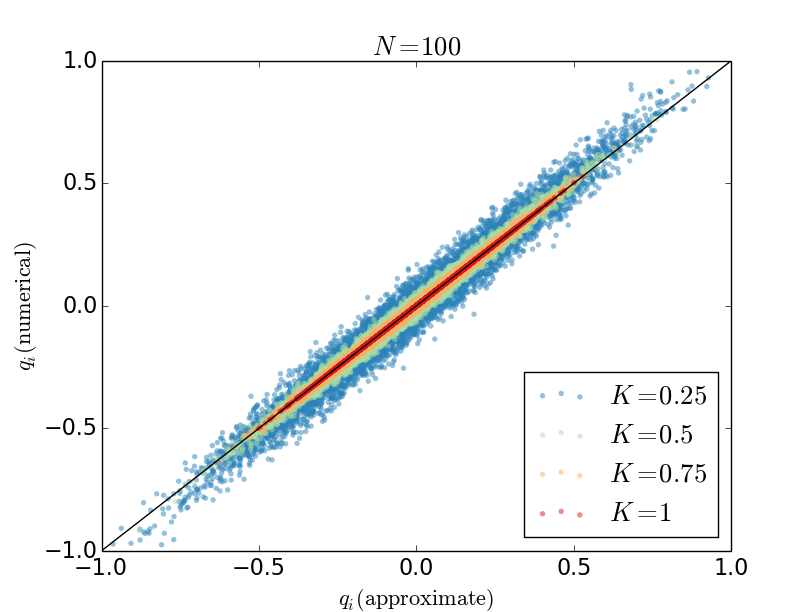}\includegraphics[width=0.5\textwidth]{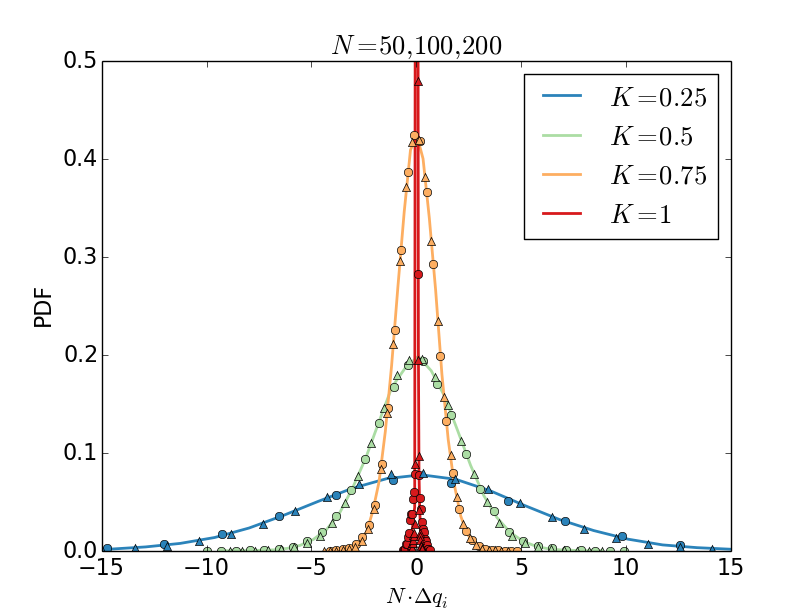}
 \caption{
We generate random Erdos-Renyi topologies of mean connectivity $K$ and size $N$, where each link $w_{i\rightarrow j}$ is an independent Gaussian variable of mean $w$ and standard deviation $\sigma w$. We set $w=1$ and $\sigma=10^{-2}$. In the absence of external flux ($J=0$), we integrate numerically the associated master equation until stationarity, and calculate $q_i = (N p_i^* -1)/\sigma$. \textbf{(Left panel)} We compare the numerical value of $q_i$ with the one given by the approximation of eq. \eqref{eq:qi-approx}, for all the nodes in the network and for $10^3$ randomly generated networks. Network size has been set to $N=100$. \textbf{(Right panel)} PDF of the error made in eq. \eqref{eq:qi-approx}, rescaled with the system size $N$, for $N=50$ (circles), $N=100$ (solid line) and $N=200$ (triangles), for different connectivities with the same colorcode as on the left panel. For each connectivity, the PDFs of the rescaled error $N\Delta q_i$ collapse into a single curve, illustrating that the error $\Delta q_i = \mathcal{O}(N^{-1})$.}
\label{fig:qi}
 \end{figure}

For a general connectivity $K$, we can neglect the implicit term in eq. \eqref{eq:qi}, leading to:
\begin{equation}
 q_i\simeq -\frac{1}{k_i} \sum_{j} A_{ij} D_{ij},
 \label{eq:qi-approx}
\end{equation}
which turns out to be a rather accurate approximation, especially for large $N$, as illustrated Fig. \ref{fig:qi} (left panel). Corrections to this formula are of $\mathcal{O}(N^{-1})$ (see Fig \ref{fig:qi}, right panel), and therefore can be neglected for large system sizes. Indeed, notice that the approximate form of eq. \eqref{eq:qi-approx} does not satisfy the constraint $\sum_i p_i^* = 1+\sigma \sum_i q_i/N=1$ in general, but the error vanishes when $N\rightarrow \infty$. 

We can introduce eq. \eqref{eq:qi-approx} into eq. \eqref{eq:Sint2}, obtaining:
\begin{equation}
 \dot S_\mathrm{int}^2 = - \frac{w \sigma^2}{N} \sum_i \frac{1}{k_i} \left( \sum_{j} A_{ij} D_{ij} \right)^2.
\end{equation}
For each $i$, $\sum_j A_{ij} D_{ij}$ is a Gaussian random variable with zero mean and variance $2 (k_i-1)$, and therefore $(\sum_j A_{ij} D_{ij})^2/k_i$ follows a chi-squared distribution of mean $2(1-k_i^{-1})\simeq 2$ and variance $8(1-k_i^{-1})^2\simeq 8$, as $k_i^{-1}$ can be neglected for large network sizes.
Finally, we have to sum over nodes $i$; in principle, one cannot apply straightforwardly the Central Limit theorem, as the elements of the sum present correlations. Think, for instance, in the simple case of a triangle network, where one has to calculate $(D_{12}+D_{13})^2 + (D_{21}+D_{23})^2 + (D_{31}+D_{32})^2$, where $D_{ij}=-D_{ji}$. However, the number of correlated elements in the sum becomes negligible with the number of uncorrelated ones for large system sizes, so one can simply neglect such a correlation. After these approximations, we find:
\begin{equation}
 P\left(\frac{\dot S_\mathrm{int}^2}{w\sigma^2}\right) \simeq \mathcal{N}\left( -2, \sqrt{\frac{8}{N}} \right).
 \label{eq:P-Sint2}
\end{equation}
Notice that $\langle \dot S_\mathrm{int}^2 \rangle$ is independent of $N$, whereas $\langle \dot S_\mathrm{int}^1 \rangle$ grows linearly with $N$ (eq. \eqref{eq:P-Sint1}). Therefore, the former can be neglected respect to the latter.

\subsection{Calculation of $\dot S_\mathrm{int}^3$}
We compute $\dot S_{\mathrm{int}}^3$ following a similar approach as for $\dot S_{\mathrm{int}}^2$. In the absence of asymmetry, $\sigma=0$, and for small external flux $J$, the stationary state is written as $p_i^* = \frac{1}{N}\left(1 + J r_i\right)$. To find $r_i$, we solve the master equation at stationarity:
\begin{equation}
 0 = \sum_{j} A_{ij} \frac{w}{N} \left(r_j-r_i\right) J + J \left(\delta_{i1}-\delta_{iN}\right)
\end{equation}
whose solution is:
\begin{equation}
 r_i = \frac{N}{w k_i}\left(\delta_{i1}-\delta_{iN} + \frac{w}{N}\sum_{j} A_{ij} r_j \right).
 \label{eq:ri}
\end{equation}
In the fully connected network ($A_{ij}=1$), the second term on the r.h.s. vanishes (since $\sum_j r_j = 0$), leading to $r_i=(\delta_{i1}-\delta_{iN})/w$. This means that all nodes have equal probabilities except for the $1$st and $N$th, that present a small unbalance $p_1 - p_N = 2 J / (Nw)$. Consequently, $w_{eq}=Nw/2$ in the fully connected network). As we did for the calculation of $\dot S_\mathrm{int}^2$, we can use this solution as an approximation for a general connectivity $K$, 
\begin{equation}
r_i \simeq \frac{N}{w k_i}\left(\delta_{i1}-\delta_{iN}\right).
\label{eq:ri-approx}
\end{equation}
As for the case of $q_i$, the approximate form above does not satisfy the constraint $\sum_i p_i^* = 1+J \sum_i r_i/N=1$ in general, but the error made vanishes when $N\rightarrow \infty$. 
Plugging eq. \eqref{eq:ri-approx} into eq. \eqref{eq:Sint3}, we obtain:
\begin{equation}
 \dot S_\mathrm{int}^3 = \sigma J \left(\frac{1}{k_1} \sum_j A_{1j} D_{1j} - \frac{1}{k_N} \sum_j A_{Nj} D_{Nj}\right).
\label{eq:Sint3-2}
 \end{equation}
Let us remind that $\sum_j A_{ij} D_{ij}$ is a Gaussian variable with zero mean and variance $2 k_i$. As we deal with Erdos-Renyi networks, we can assume that $k_1 \approx k_N \approx K N$; furthermore, if $A_{1N}=0$, there is no correlation between the first and the second contribution in eq. \eqref{eq:Sint3-2}; otherwise, a small correlation may exist for large network sizes $N$, but this can be neglected. With these approximations, we finally obtain that
\begin{equation}
P\left(\frac{\dot S_\mathrm{int}^3}{\sigma J}\right) \simeq \mathcal{N}\left(0,\frac{2}{\sqrt{K N}} \right).
\label{eq:P-Sint3}
\end{equation}
Observe that, $\langle \dot S_\mathrm{int}^{3}\rangle$ is independent of $N$, and therefore this contribution can be neglected respect to the one given by $\dot S_\mathrm{int}^1$.

\begin{figure}[h]
 \centering
 \includegraphics[width=0.5\textwidth]{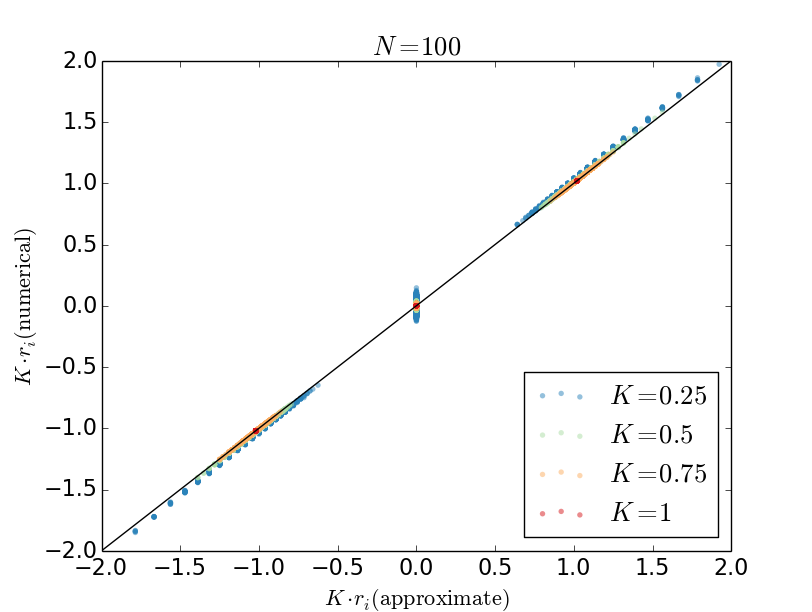}\includegraphics[width=0.5\textwidth]{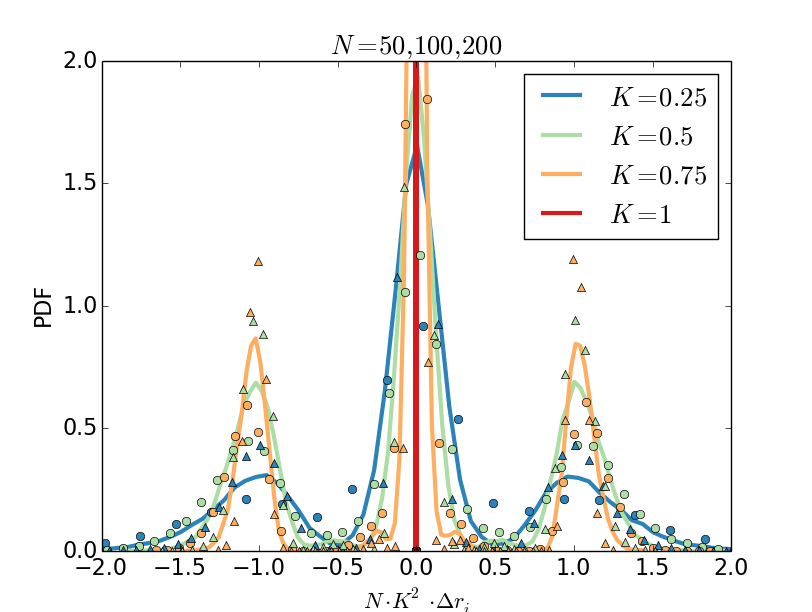}
\caption{
We generate random Erdos-Renyi topologies of mean connectivity $K$ and size $N$, where each link $w_{i\rightarrow j}=w=1$ (i.e. no asymmetry). The external flux is set to $J=10^{-2}$. We integrate numerically the associated master equation until stationarity, and calculate $r_i = (N p_i^* -1)/J$. \textbf{(Left panel)} We compare the numerical value of $r_i$ with the one given by the approximation of eq. \eqref{eq:ri-approx}, for all the nodes in the network and for $10^3$ randomly generated networks. Network size has been set to $N=100$. \textbf{(Right panel)} PDF of the error made in eq. \eqref{eq:ri-approx}, rescaled with the system size, for $N=50$ (circles), $N=100$ (solid line) and $N=200$ (triangles), for different connectivities with the same colorcode as on the left panel. In this case, we were able to capture also the scaling with the connectivity $K$. When rescaled properly, the location of the peaks approximately collapse, illustrating that the error made is $\Delta r_i = \mathcal{O}(K^{-2} N^{-1})$.}
\label{fig:ri}
\end{figure}

\subsection{Distribution of $\dot S_\mathrm{int}$}
We can obtain the PDF of $\dot S_\mathrm{int}$ assuming that, approximately, $\dot S_1$, $\dot S_2$ and $\dot S_3$ are independent random variables, so from Eqs. \eqref{eq:P-Sint1}, \eqref{eq:P-Sint2} and \eqref{eq:P-Sint3} we find:
\begin{eqnarray}
 P\left(\dot S_\mathrm{int}=\dot S_1 + \dot S_2 + \dot S_3 \right) \simeq 
\mathcal{N}\left(
(K (N-1)-2) w \sigma^2 , \sqrt{ \left(4 K\left(1-\frac{1}{N}\right)  + \frac{8}{N}\right)\left(w \sigma^2\right)^2 + \frac{4}{K N} \left(\sigma J\right)^2 }
\right) \\
\xrightarrow{\quad N\gg 1 \quad } \mathcal{N}\left((K N - (2+K)) w \sigma^2  , 2\sqrt{K} w \sigma^2  \right),
\label{eq:P-Sint}
\end{eqnarray}
where in the last expression we have considered the leading contribution for large system size $N$.

\section{Distribution of $w_\mathrm{eq}$}
\begin{figure}[h]
 \centering
 \includegraphics[width=0.6\textwidth]{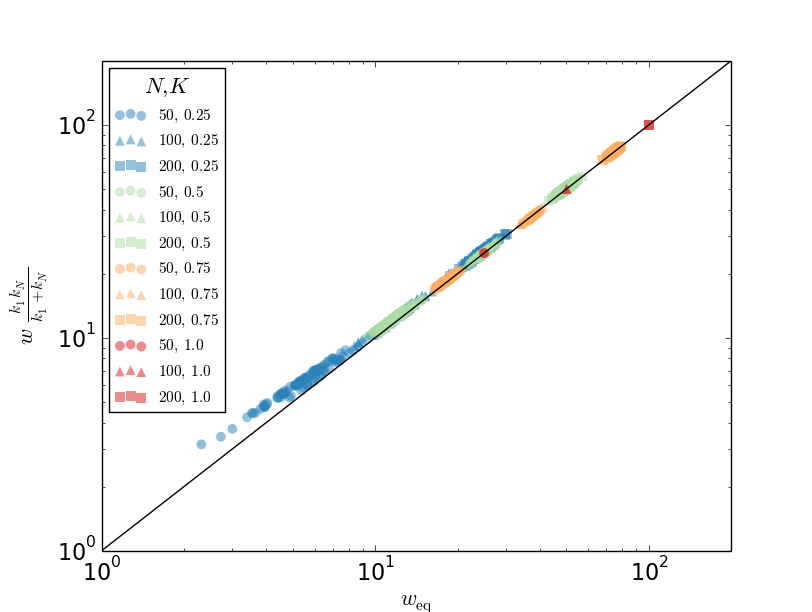}
 \caption{Comparison between $w_\mathrm{eq}$ with the approximation given by eq. \eqref{eq:weq} in symmetric master-equation systems with networks of size $N=50,100,200$ and connectivity $K=0.25,0.5,0.75,1$. All existing links in the network has a rate value $w=1$. For each network, $w_\mathrm{eq}$ has been obtained by integrating numerically the master equation with an additional external flux $J=0.01$, and then calculating $w_\mathrm{eq}=(p_1^*-p_N^*)/J$.
 }
 \label{fig:weq0}
\end{figure}

The equivalent transition rate is defined close to equilibrium as $w_\mathrm{eq} = \frac{J}{p_1^* - p_N^*}$ when $\sigma=0$. Introducing the explicit form of $p_i^*\simeq(1+J r_i)/N$, where $r_i$ has the form of eq. \eqref{eq:ri-approx} in large networks, we obtain:
\begin{equation}
 w_\mathrm{eq} \simeq w \frac{k_1 k_N}{k_1 + k_N}.
 \label{eq:weq}
\end{equation}
Consequently, variability in $w_\mathrm{eq}$ comes, essentially, from the variability in the degrees of $k_1$ and $k_N$. We check numerically the validity of eq. \eqref{eq:weq}. To this end, we generate master-equation systems described by random networks of size $N$ and connectivity $K$, and for each one we compare the corresponding value of $w_{eq}$ (obtained integrating numerically the master equation) with the one given by eq. \eqref{eq:weq}. The result is plotted in Fig. \ref{fig:weq0} in log-log scale, showing a good agreement, especially for large network sizes and higher connectivities.

\begin{figure}[h]
 \centering
\includegraphics[width=0.5\textwidth]{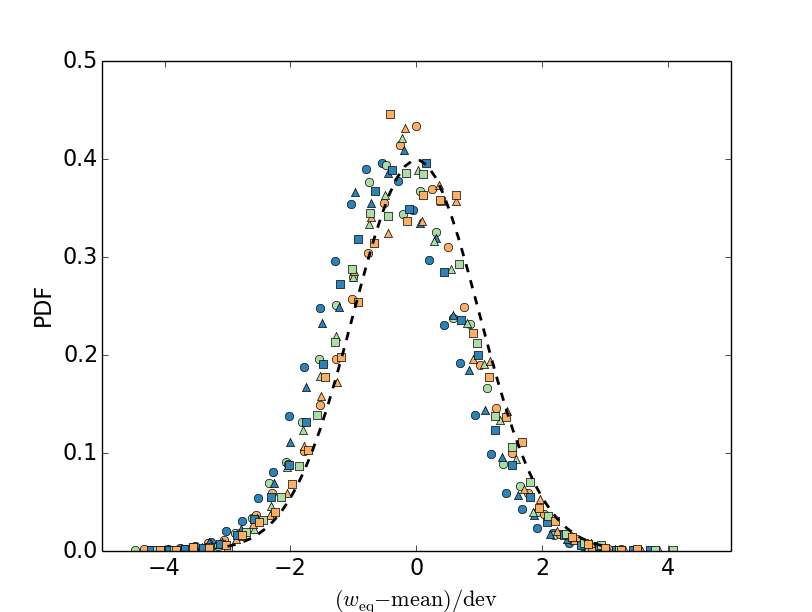}
\caption{Collapse of the PDF for 9 datasets of $w_\mathrm{eq}$, for $K=0.25$ (blue), $K=0.5$ (green) and $K=0.75$ (yellow), and for each one, for $N=50$ (circles), $N=100$ (triangles) and $N=200$ (squares)  (same datasets of Fig. \ref{fig:weq0}), obtained by subtracting the corresponding mean $NKw/2$ and dividing by the standard deviation $NK(1-K)w/8$ on each data (Eqs. \eqref{eq:weq-mean} and \eqref{eq:weq-var} for the left panel and Eqs. \eqref{eq:weqinv-mean} and \eqref{eq:weqinv-var} for the right one). Dashed lines represent a Gaussian distribution of zero mean and unit variance. Deviations stem from finite size corrections, and indeed PDFs for larger system sizes fit better the Gaussian distribution.
}
\label{fig:weq}
\end{figure}

The distribution of $w_\mathrm{eq}$ must be computed from the degree distribution of an Erdos-Renyi network of connectivity $K$, which is given by the binomial distribution 
$p(k) = {N \choose k} K^k (1-K)^{N-k}$.  For large $N$, the binomial distribution behaves as a Gaussian distribution with mean $NK$ and variance $NK(1-K)$, and neglecting the correlation between $k_1$ and $k_N$ we can write:
\begin{equation}
 P(w_\mathrm{eq})\simeq\int_{-\infty}^\infty\int_{-\infty}^\infty  \delta\left(w_\mathrm{eq}-w\frac{k_1 k_N}{k_1+k_N}\right) P(k_1) P(k_N) dk_1 dk_N,
\end{equation}
where the limits of the integral have been extended the whole real axis and $P(k)$ represents a Gaussian distribution $\mathcal{N}(NK, \sqrt{NK(1-K)})$. 
The non-linear dependency of $w_{eq}$ on $k_1$ and $k_N$ hinders the calculation of $P(w_{eq})$. However, as the Gaussian distributions becomes narrower (in relation to their mean) when $N$ increases, we can calculate the PDF in the limit of large $N$ using a saddle-point approximation. First, we write the delta function in terms of its Fourier representation:
\begin{equation}
  P(w_\mathrm{eq}) = \int_{-\infty}^\infty\int_{-\infty}^\infty \int_{-\infty}^\infty  \frac{1}{2\pi} \exp\left[-i \alpha \left(w_\mathrm{eq}-w \frac{k_1 k_N}{k_1 + k_N}\right) \right] P(k_1) P(k_N) dk_1 dk_N d\alpha.
\end{equation}
Inserting the explicit form of $P(k)$, using the change of variables: $x_1=\dfrac{k_1}{K N} - 1$, $x_N = \dfrac{k_N}{K N} - 1$, and introducing the intensive variable $\hat{w}_\mathrm{eq}= \dfrac{w_\mathrm{eq}}{N}$, we obtain:
\begin{eqnarray}
   P(w_\mathrm{eq}) &=& \frac{N K}{(2\pi)^2(1-K)} \int_{-\infty}^\infty\int_{-\infty}^\infty \int_{-\infty}^\infty  \exp\left[ N\left(-i\alpha \left( \hat{w}_\mathrm{eq} - w K \frac{(1+x_1)(1+x_N)}{2+x_1+x_N} \right) - \frac{K}{2(1-K)}(x_1^2+x_N^2)\right)\right] dx_1 dx_N d\alpha\nonumber\\
   &\equiv& \frac{N K}{(2\pi)^2(1-K)} \int_{-\infty}^\infty\int_{-\infty}^\infty \int_{-\infty}^\infty  \exp\left[ N F(x_1, x_N, \alpha) \right] dx_1 dx_N d\alpha.
\end{eqnarray}
The saddle-node approximation states that the integral $\int d^dz \exp[NF(\vec{z})] \simeq (2\pi/N)^{d/2}(-\det H(F)|_{\vec{z}^*})^{-1/2} \exp[N f(\vec{z}^*)]$ in the limit of large $N$, where $\vec{z}^*$ is the stationary point of $F$, $d$ its dimension and $H$ the Hessian matrix of $F$. Imposing $\partial _{x_1} F = \partial _{x_N} F = \partial _{\alpha} F = 0$, we can find that in our case the stationary point is located at $x_1^* = x_N^* = 2 \frac{\hat{w}_\mathrm{eq}}{KN} - 1$, $\alpha^* = 4 i \frac{k w - 2 \hat{w}_\mathrm{eq}}{k (1-k) w^2}$. After some calculations, and written in terms of the original variable $w_\mathrm{eq}$, we find:
\begin{equation}
P(w_\mathrm{eq}) \simeq \frac{1}{\sqrt{ 2 \pi  \frac{N K (1-K)}{8} w^2 \left(3-2 \frac{w K N}{2 w_\mathrm{eq}} \right) }} \exp\left[-\frac{\left(w_\mathrm{eq}-\frac{w K N}{2} \right)^2}{2 \frac{N K (1-K)}{8} w^2}\right].
\end{equation}
As the PDF concentrates around $w_\mathrm{eq}\simeq \frac{wKN}{2}$ when $N$ increases, $\left(3-2 \frac{w K N}{2 w_\mathrm{eq}} \right)\simeq 1$. Therefore, we finally obtain that
 \begin{equation}
P\left(\frac{w_\mathrm{eq}}{w}\right) \simeq \mathcal{N}\left(\frac{NK}{2},\sqrt{\frac{1}{8} N K (1-K)} \right).
\label{eq:pweq-gaussian}
\end{equation}
We can collapse data of $w_\mathrm{eq}$ for different ensembles into a single PDF by substracting the mean and rescaling by the standard deviation as obtainable from eq. \eqref{eq:pweq-gaussian}. Results are shown in Fig. \ref{fig:weq}. Deviations from a perfect collapse (that are higher for smaller networks) must stem from finite size corrections to the scaling formulas (see Section \ref{sec:weq-moments}). 

\subsection{Calculation of $\langle f(w_\mathrm{eq})\rangle$}
\label{sec:weq-moments}
We can calculate the expected value of some general function $f(w_\mathrm{eq})$ using the asymptotic form of $P(w_\mathrm{eq})$, Eq. \eqref{eq:pweq-gaussian}. However, it can be useful to calculate next-to-leading order terms as it follows. We first write the expected value as
\begin{equation}
  \langle f(w_\mathrm{eq})\rangle = \int_{-\infty}^\infty\int_{-\infty}^\infty  f\left(w \frac{k_1 k_N}{k_1 + k_N}\right) P(k_1) P(k_N) dk_1 dk_N = \sum_{n,m=0}^\infty \frac{\mu_{2n} \mu_{2m}}{(2n)!(2m)!} \left. \frac{\partial f}{\partial k_1^{2n} \partial k_N^{2m}}\right|_{k_1=k_N=KN},
\end{equation}
where $\mu_i$ is the $i$-th central moment of a Gaussian distribution of mean $KN$ and variance $NK(1-K)$, and them simply truncate the series at the desired order.  Using this expression, it is straightforward to calculate $\langle w_\mathrm{eq} \rangle$ expanding $\frac{k_1 k_N}{k_1+k_N}$ up to the the forth moment:
\begin{equation}
\label{eq:weq-mean}
 \langle w_\mathrm{eq} \rangle = w \left(\frac{KN}{2} - \frac{1-K}{4}\right) -\frac{(1-K)^2}{8KN} + \mathcal{O}(N^{-2})
\end{equation}
and its variance $\langle\langle w_\mathrm{eq}^2 \rangle\rangle = \langle w_\mathrm{eq}^2 \rangle - \langle w_\mathrm{eq} \rangle^2$ from the expansion of $\left(\frac{k_1 k_N}{k_1+k_N}\right)^2$:
\begin{equation}
\label{eq:weq-var}
 \langle\langle w_\mathrm{eq}^2 \rangle\rangle = \frac{w^2}{8}\left(NK(1-K)+2(1-K)^2\right)+ \mathcal{O}(N^{-1}).
\end{equation}
As expected, the leading orders agree with the mean and variance of $P(w_\mathrm{eq})$, eq. \eqref{eq:pweq-gaussian}.
Similarly, it is also useful to compute the mean and variance of $w_\mathrm{eq}^{-1}$ using the similar procedure; the result is:
\begin{eqnarray}
\label{eq:weqinv-mean}
\langle w_\mathrm{eq}^{-1} \rangle &=& \frac{2}{w K N} \left( 1 + \frac{1-K}{K N} \right) + \mathcal{O}(N^{-3}) \\
\label{eq:weqinv-var}
\langle\langle (w_\mathrm{eq}^{-1})^2\rangle\rangle &=&  \frac{2 (1-K) }{w^2 (K N)^3} \left(1+20\frac{1-K}{KN}\right) + \mathcal{O}(N^{-5}).
\end{eqnarray}
Let us notice that the leading term of $\langle w_\mathrm{eq}^{-1} \rangle$ is the one used in the bottom panel of Fig. \ref{fig:3}.

\section{Distribution of $\epsilon_\mathrm{eq}$}
To compute the unbalance parameter $\epsilon_\mathrm{eq}$, we set the external current to $J=0$. In this setting, and for small asymmetry values, $\sigma\ll 1$, the stationary probability can be written as $p_i^*=1/N(1+\sigma q_i)$, so $\epsilon_{\mathrm{eq}}$ can be computed as:
\begin{equation}
 \epsilon_\mathrm{eq}=\frac{p_1^*-p_N^*}{\sigma} = \frac{q_1 - q_N}{N}.
\end{equation}
Introducing the approximation given by eq. \eqref{eq:qi-approx} into the equation above, one finds the explicit formula:
\begin{equation}
 \epsilon_\mathrm{eq}=\frac{1}{N}\left(\frac{1}{k_N} \sum_j A_{Nj} D_{Nj} - \frac{1}{k_1} \sum_j A_{1j} D_{1j}\right).
 \label{eq:epsilon-eq-1}
\end{equation}
At this step, we neglect the heterogeneity in the connectivity degree and set $k_1\simeq k_N \simeq K N$. 
Reminding that $D_{ij}=-D_{ji}$ are independent Gaussian variables for each pair of links with zero mean and variance 2, for large networks the Central Limit Theorem gives:
\begin{equation}
 P(\epsilon_{\mathrm{eq}})\simeq\mathcal{N}\left(0, \frac{2}{N \sqrt{K N}}\right),
 \label{eq:Pepsilon}
\end{equation}
where we have neglected a small correction arising if nodes $1$ and $N$ are connected (i.e. when $A_{1N}=1$); in such a case the first and second term in eq. \eqref{eq:epsilon-eq-1} would not be completely uncorrelated, as $D_{1N}=-D_{N1}$. In any case, this correction is negligible for large N.

\begin{figure}[h]
 \centering
 \includegraphics[width=0.6\textwidth]{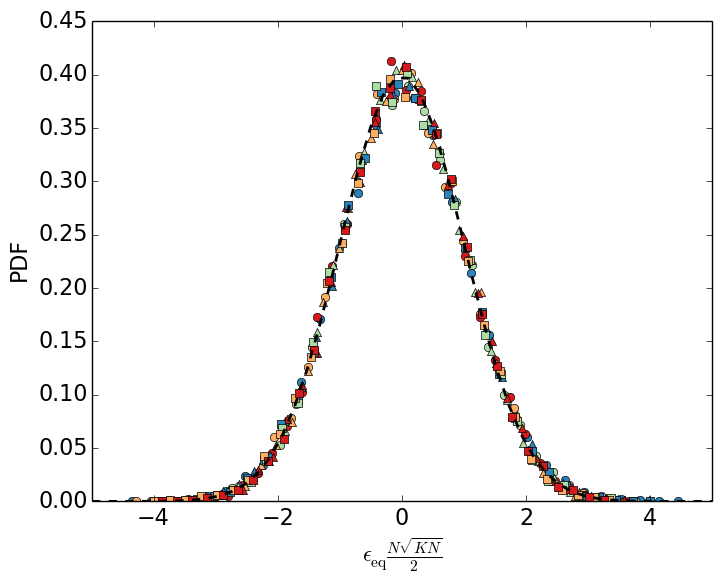}
\caption{
Collapse of different $P(\epsilon_\mathrm{eq})$ for asymmetric master-equation systems in ensembles of $10^4$ networks of size $N=50$ (circles), $N=100$ (triangles) and $N=200$ (squares), and for each one, for connectivities $K=0.25$ (blue), $K=0.5$ (green) and $K=0.75$ (yellow). Link weights in the network are independent Gaussian variables of mean $w=1$ and standard deviation $w\sigma=0.01$. For each network, $\epsilon_\mathrm{eq}$ has been obtained by integrating numerically the master equation with the external flux $J=0$, and then calculating $\epsilon_\mathrm{eq}=(p_1^*-p_N^*)/\sigma$. Dark dashed line represents the Gaussian distribution of zero mean and unit variance. 
The collapse has been obtained by dividing each value of $\epsilon_\mathrm{eq}$ by its corresponding standard deviation obtained analytically, $2/(N\sqrt{K N})$.
}
\label{fig:weq2}
\end{figure}

\section{Deviations from Joule's law}
Close to equilibrium, we can write the total entropy production for asymmetric systems at stationarity as $\dot S^* = \dot S_\mathcal{J} + N \epsilon_\mathrm{eq} \sigma J + \dot S_\mathrm{int}$ (see main text). Therefore, deviations from Joule's law, $\delta \dot S_\mathcal{J} = \dot S^* - \dot S_\mathcal{J}$ can be obtained combining the Gaussian distributions of $\epsilon_\mathrm{eq}$ (eq. \eqref{eq:Pepsilon}) and $\dot S_\mathrm{int}$ (eq. \eqref{eq:P-Sint}): 
\begin{equation}
P\left(\delta \dot S_\mathcal{J} = \dot S_\mathrm{int} +  N\epsilon_\mathrm{eq} \sigma J\right) \simeq 
\mathcal{N}\left(
(N K-(2+K)) w \sigma^2 , \sqrt{4 K \left(w \sigma^2\right)^2 + \frac{4}{K N}  \left(\sigma J\right)^2 } 
\right)
\xrightarrow{\quad NK\gg 1 \quad } \mathcal{N}\left(N K w \sigma^2  , 2\sqrt{K} w \sigma^2  \right).
\end{equation}
We can see that, for large systems, the contribution given by $\epsilon_\mathrm{eq}$ is negligible, obtaining the same result as in eq. \eqref{eq:P-Sint}.

\end{document}